\documentclass[reprint,amsmath,amssymb,aps,prapplied,superscriptaddress,longbibliography]{revtex4-2}

\usepackage{graphicx}
\usepackage{dcolumn}
\usepackage{bm}
\usepackage{upgreek}
\usepackage{xcolor}
\usepackage{overpic}
\usepackage{multirow}
\usepackage{hyperref}
\usepackage{gensymb}

\usepackage{color}

\usepackage[normalem]{ulem}

\newcommand{\figref}[1]{Fig.~\ref{fig:#1}}
\renewcommand{\Re}{\operatorname{Re}}
\renewcommand{\Im}{\operatorname{Im}}
\renewcommand{\vec}[1]{\mathbf{#1}}

\begin{document}

\title{Robustness of Bound States in the Continuum in Bilayer Structures against Symmetry Breaking}

\author{Kliment V. Semushev}
\thanks{These authors contributed equally.}
\affiliation{Qingdao Innovation and Development Center of Harbin Engineering University, 266500, Qingdao, China}
\affiliation{School of Physics and Engineering, ITMO University, 197101, St. Petersburg, Russia}

\author{Zilong Zhao}
\thanks{These authors contributed equally.}
\affiliation{Qingdao Innovation and Development Center of Harbin Engineering University, 266500, Qingdao, China}

\author{Alexey Proskurin}
\affiliation{Qingdao Innovation and Development Center of Harbin Engineering University, 266500, Qingdao, China}
\affiliation{School of Physics and Engineering, ITMO University, 197101, St. Petersburg, Russia}

\author{Mingzhao Song}
\email{kevinsmz@foxmail.com}
\affiliation{Qingdao Innovation and Development Center of Harbin Engineering University, 266500, Qingdao, China}

\author{Xinrui Liu}
\affiliation{Qingdao Innovation and Development Center of Harbin Engineering University, 266500, Qingdao, China}

\author{Mikhail V. Rybin}
\affiliation{School of Physics and Engineering, ITMO University, 197101, St. Petersburg, Russia}
\affiliation{Ioffe Institute, 194021, St. Petersburg, Russia}

\author{Ekaterina E. Maslova}
\email{ekaterina.maslova@metalab.ifmo.ru}
\affiliation{Qingdao Innovation and Development Center of Harbin Engineering University, 266500, Qingdao, China}
\affiliation{School of Physics and Engineering, ITMO University, 197101, St. Petersburg, Russia}

\author{Andrey A. Bogdanov}
\email{bogdan.taurus@gmail.com}
\affiliation{Qingdao Innovation and Development Center of Harbin Engineering University, 266500, Qingdao, China}
\affiliation{School of Physics and Engineering, ITMO University, 197101, St. Petersburg, Russia}

\date{\today}

\begin{abstract}
We investigate the robustness of bound states in the continuum (BICs) in a bilayer dielectric rod array against geometric and material perturbations. Our analysis focuses on both symmetry-protected and Fabry-Pérot BICs, examining their transformation into quasi-BICs under three structural modifications: (i) in-plane displacement of one layer, which breaks the C$_2$ symmetry of the system; (ii) introduction of material losses that break time-reversal symmetry; and (iii) variation in the interlayer distance, which preserves structural symmetry. In particular, we demonstrate that material losses inevitably induce radiation in Fabry-Pérot BICs via second-order perturbation processes, converting them into quasi-BICs, while symmetry-protected BICs remain non-radiative. We further show that, despite the inherent instability of BICs under symmetry-breaking effects, their resilience can be significantly enhanced through proper design. Both Fabry-Pérot and symmetry-protected BICs exhibit exponentially weak sensitivity to C$_2$-breaking perturbations as the interlayer distance increases. Finally, we show that additional FP-BICs emerge under oblique incidence, originating from the interference of two high-Q quasi-BICs near the symmetry-protected ones. Our findings pave the way for the development of BIC-based photonic devices with improved robustness against fabrication imperfections, environmental variations, and material losses.
\end{abstract}

\maketitle


\section{Introduction} \label{Introduction}

Bound states in the continuum (BICs) are non-radiating states that exist within the continuum spectrum of radiating modes in the surrounding space~\cite{vonNeumann1929,koshelev2020engineering}. 
The radiative $Q$ factor of BICs diverges; however, in real systems, BICs transform into quasi-BICs (qBICs) exhibiting a finite radiative lifetime and a finite radiative $Q$ factor~\cite{koshelev2020engineering, hsu2013observation}. The radiative $Q$ factor of quasi-BICs (qBICs) can be precisely tuned controlling symmetry breaking in the structure or by adjusting the angle of incidence~\cite{KuhneWangWeberKuhnerMaierTittl,MaslovaRybinBogdanovSadrieva}. By adjusting the radiative $Q$ factor of quasi-BICs, one can achieve the critical coupling regime, in which the radiative and non-radiative losses are perfectly balanced. This results in maximum field enhancement, which is important for many applications, including sensing~\cite{Romano2018,Srivastava2019,Maksimov2022}, lasing~\cite{Kodigala2017,Yu2021,Hwang2021}, nonlinear optics~\cite{carletti2018giant,carletti2019high,koshelev2020subwavelength}, and polaritonics~\cite{Kravtsov2020nonlinear,Berghuis2023room,Luo2025room,Koshelev2018strong,Maggiolini2023strongly,Weber2023intrinsic}.

In periodic photonic structures, BICs are typically classified into (i) symmetry-protected BICs~\cite{ni2016tunable,cong2019symmetry} and (ii) accidental BICs~\cite{plotnik2011experimental,koshelev2018asymmetric, wu2024observation}. The coupling of symmetry-protected BICs to open scattering channels is forbidden due to the selection rules, i.e. 
mismatch between the symmetry of the mode and the radiating waves. In contrast, to avoid coupling of accidental BICs to open scattering channels, 
geometric or material parameters of the system should be tuned precisely. The simplest example of an accidental BIC is a \textit{Fabry-Pérot} BIC (FP-BIC), which is formed in an effective 1D system composed of two identical resonant scatterers, i.e.~mirrors that perfectly reflect waves at the resonance
frequency. FP-BICs appear when the distance between the scatterers satisfies the Fabry-Pérot quantization condition at the resonance
frequency of the mirrors. Resonant mirrors can be formed by gratings, metasurfaces, or even the upper and lower interfaces of a photonic crystal slab~\cite{suh2005displacement,hsu2013bloch,anguiano2014spectra,bikbaev2017optical,wu2022tailoring,ovcharenko2020bound,gao2016formation}. 

FP-BICs have been extensively studied in a variety of photonic systems~\cite{liu2009resonance,Luo2022wavy,Nabol2022fabry,Rao2025manipulation,Mai2025relationship,Ni2024three,Alagappan2024fabry,Ndangali2010electromagnetic,bulgakovPropagatingBoundStates2018,doskolovichResonantPropertiesComposite2019,liBoundStatesContinuum2016,shiDoublelayerSymmetricGratings2022}.
The symmetry of metasurfaces or gratings was demonstrated to play a crucial role in the formation of FP-BICs. Ndangali and Shabanov were the first to provide a detailed analysis of FP-BICs in bilayer gratings and their robustness to symmetry breaking, supporting the analysis with a deep mathematical foundation \cite{Ndangali2010electromagnetic}. Moreover, Nabol et al.~\cite{Nabol2022fabry} demonstrated that an anisotropic photonic crystal with two anisotropic defect layers can support FP-BICs, with an analytic solution explaining spectral features such as avoided crossings and resonance collapses, which enables
the design of microcavities with controllable $Q$ factors. The tuning of FP-BICs is also influenced by parameters such as the distance between layers and asymmetry in periodic structures~\cite{Rao2025manipulation,Mai2025relationship}. In some cases, symmetry-breaking effects prevent the formation of FP-BICs, even when total reflection conditions are met, illustrating that perfect total reflection does not always guarantee the emergence of FP-BICs~\cite{Mai2025relationship}. The sensitivity of FP-BICs to system parameters can be harnessed for dynamic control of their $Q$ factor through adjustments such as 
phase detuning and structural modifications~\cite{Rao2025manipulation}. These findings emphasize that symmetry is essential in the formation and manipulation of FP-BICs, {and} 
the influence of symmetry on these phenomena remains {under active investigation}
~\cite{Ni2024three}.

\begin{figure}[t]
   \includegraphics[width=1\linewidth]{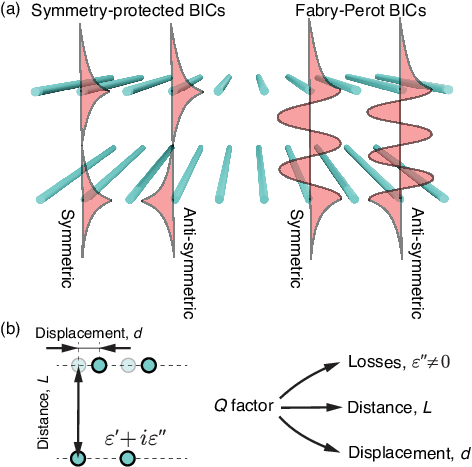}
  \caption{(a) Classification of BICs in bilayer structure. (b) Schematic view of the considered resonator and the parameters affecting the $Q$ factor.}
   \label{fig:Structure}
\end{figure}

In this paper, we present a comprehensive analysis of the $Q$ factor for both symmetry-protected and FP-BICs in a bilayer periodic array of infinitely long dielectric rods [\figref{Structure}(a)]. We analyze 
the effect of material losses (T-symmetry breaking), 
variations in the interlayer distance, and in-plane displacement of one layer (C\textsubscript{2}-symmetry breaking) [\figref{Structure}(b)]. We analytically reveal how material losses affect FP-BICs and show that absorption not only leads to dissipative losses, but also induces additional radiative leakage as a second-order perturbation effect, thereby transforming FP-BICs into quasi-FP-BICs with a finite radiative lifetime. In contrast, symmetry-protected BICs remain robust against material losses. Furthermore, we demonstrate that for a fixed material loss, the $Q$ factor of symmetry-protected BICs is largely insensitive to variations in the interlayer distance, whereas the $Q$ factor of FP-BICs increases linearly with the distance between the layers. 
We also demonstrate that despite the inherent instability of BICs under C\textsubscript{2}-symmetry-breaking, their resilience can be significantly enhanced through design optimization. Specifically, both FP and symmetry-protected BICs exhibit exponentially weak sensitivity to C\textsubscript{2}-breaking perturbations as the interlayer distance increases.
\begin{figure}[t]
    \includegraphics[width=1\linewidth]{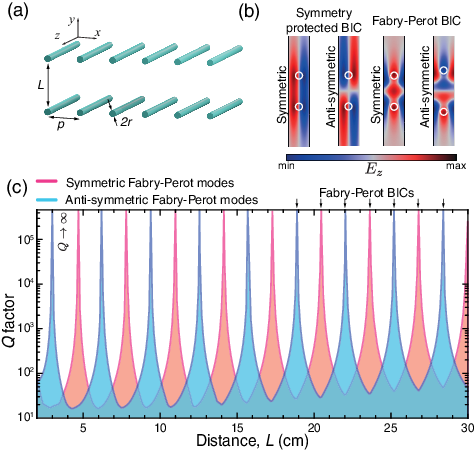}
    \caption{Bound states in the continuum in the bilayer structure of infinite dielectric rods. (a) Schematic view of the photonic structure. (b) Distribution of the $E_\mathrm{z}$ electric field of the Fabry-Pérot and symmetry-protected BICs at the $\Gamma$-point. (c) Dependence of $Q$ factors of different BICs on the distance between {the} layers of dielectric rods $L$.
    }
    \label{fig:Qh}
\end{figure}
\section{Theoretical background}
\subsection{Infinite array of dielectric rods}\label{sec:Bloch}

We consider a metasurface consisting of two identical periodic arrays of infinitely long parallel dielectric rods [\figref{Qh}(a)]. The radius of the rods is $r = 0.5$~cm and the period is $p = 3$~cm. Thus, the unit cell of the structure contains two rods placed at a distance $L$ from each other. The permittivity of the rods is $\varepsilon = 2.1$, which corresponds to low-loss polymers, such as Teflon in the frequency range of 0.1-10~GHz. A similar structure has been studied in detail in Ref. \cite{bulgakovBlochBoundStates2014}. It supports two types of BIC{s}: (i) FP-BICs~\cite{suh2005displacement,shubin2023twin,rezzouk2024fabry,Rao2025manipulation} and (ii) symmetry-protected BICs~\cite{Pankin_Maksimov_Chen_Timofeev_2020,PhysRevB.89.165111,Foley:15}. We should emphasize that symmetry-protected-BICs (SP-BICs) arise in single-layer arrays due to symmetry mismatch with the radiation continuum, whereas FP-BICs occur only in bilayer configurations when the interlayer distance satisfies the Fabry-Pérot condition. For more details on the connection between Bloch{'s} theorem and different types of BICs, see Appendix~\ref{app:Bloch}. We limit our analysis to the case of TE-polarized modes [$\vec E=(0,0,E_z)$] assuming $k_z=0$.

Since symmetry-protected BICs exist independently in each single-layer structure, in the bilayer configuration, they couple via the near field, forming symmetric and anti-symmetric symmetry-protected BICs [see Fig.~\ref{fig:Qh}(b); we {thoroughly discuss the mode symmetry} 
in Appendix~\ref{app:symmetry}]. Varying the interlayer distance $L$ preserves the mirror symmetry of the structure and does not destroy these BICs. In contrast, FP-BICs emerge only at specific distances $L$ that satisfy Fabry-Pérot quantization conditions at frequencies near the resonan{ce} 
frequency of the single-layer metasurface. Figure~\ref{fig:Qh}(c) shows the dependence of the $Q$ factor of the FP-BICs on the interlayer distance $L$: the $Q$ factor varies from infinity to several tens, illustrating the extreme sensitivity of FP-BICs to changes in $L$. As variations in $L$ preserve the $xy$-mirror symmetry, both FP-BICs and symmetry-protected BICs can be considered as symmetric or anti-symmetric with respect to the transformation $y \rightarrow -y$.

\subsection{Coupled mode theory}\label{ss:cmt}

\subsubsection{Single-layer structure}\label{sec:single-layer-structure}

\begin{figure}[t!]
    \includegraphics[width=1\linewidth]{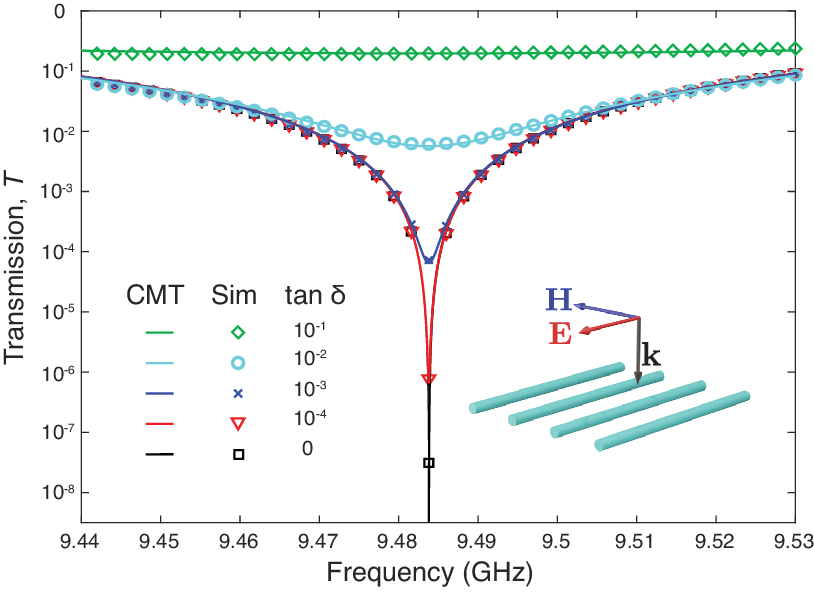}
    \caption{Comparison of numerically simulated (markers) and analytical (solid lines) transmission spectra of a single-layer structure for different losses: $\tan\delta=0$ (black), $\tan\delta=10^{-4}$ (red), $\tan\delta=10^{-3}$ (blue), $\tan\delta=10^{-2}$ (cyan), {and} $\tan\delta=10^{-1}$ (green). 
    {Inset} shows single-layer structure and {the propagation} direction of {the} inciden{t} 
    field.}
    \label{fig:fig3}
\end{figure}

FP-BICs are formed similarly to resonances in a Fabry-Pérot resonator, where the field is confined between two mirrors. Therefore, to analytically describe FP-BICs in our bilayer configuration, first, we need to study the single-layer structure and analyze its transmission spectrum. The markers in Fig.~\ref{fig:fig3} show the
transmission spectra of the single-layer structure numerically calculated using COMSOL Multiphysics for a normally incident TE-polarized plane wave for different material losses $\tan\delta = \Im \varepsilon/\Re \varepsilon$ varying from 0 to 10$^{-1}$. The structure exhibits a pronounced resonant behavior near the frequency 9.48~GHz, where the transmission spectrum has a sharp minimum. For the lossless case, the transmission becomes zero, and the structure behaves as a perfect mirror; when we introduce material losses, the structure becomes partially transparent.

The transmission spectra for a single layer can be well described using the temporal coupled mode theory (CMT)~\cite{Fan_Suh_Joannopoulos_2003}. Indeed, in terms of CMT, our structure is a single-mode resonator coupled to two ports corresponding to plane waves above and beneath the layer. The amplitude reflection ($r$) and transmission ($t$) coefficients can be written as follows (see Appendix~\ref{app:cmt} for derivation details):
\begin{equation}\label{eq:refl}
    r=\frac{-\gamma_r}{i(\omega_0-\omega)+\gamma_r+\gamma_a},
\end{equation}
\begin{equation}\label{eq:tr}
    t=\frac{i(\omega_0-\omega)+\gamma_a}{i(\omega_0-\omega)+\gamma_r+\gamma_a}.
\end{equation}
Here, $\omega_0$ is the resonant frequency, and $\gamma_r$ and $\gamma_a$ 
are the radiative and non-radiative inverse lifetimes of the resonance mode, respectively. Since CMT is a phenomenological framework, the values of $\omega_0$, $\gamma_r$, and $\gamma_a$ are obtained from the numerical solution of the eigenvalue problem in COMSOL Multiphysics. The main challenge lies in distinguishing $\gamma_r$ and $\gamma_a$ from the total decay rate $\gamma_t = \gamma_r + \gamma_a$ calculated numerically. Noting that $\gamma_t = \gamma_r$ when $\tan \delta = 0$, the absorption losses $\gamma_a$ can be determined as a function of the material loss tangent $\tan \delta$ by the relation $\gamma_a(\tan\delta) = \gamma_t - \gamma_r$. Our analysis shows that the absorption losses are almost proportional to the material loss tangent $\gamma_a \approx 1.2 \tan\delta$ {within} 
a wide range of material losses.

The solid lines in Fig.~\ref{fig:fig3} indicate the transmission spectra $T = |t|^2$ calculated using CMT [see Eq.~\eqref{eq:tr}], which are in good agreement with
the results of numerical simulations. Furthermore, Eq.~\eqref{eq:tr} implies that at the resonant frequency, the transmission behaves as
\begin{equation}\label{eq:part_transparencey}
T = |t|^2 = \frac{\gamma_a^2}{(\gamma_a + \gamma_r)^2}.
\end{equation}
From this relation, one can immediately conclude that FP-BICs are destroyed by material losses. Indeed, when $\gamma_a \neq 0$, the transmission becomes nonzero at the resonance frequency, indicating that the FP-BIC transforms into a quasi-FP-BIC with finite radiative losses.

Let us note that the single-layer structure supports an SP-BIC, but it does not manifest in the transmission or reflection spectra at normal incidence. At oblique incidence, however, the SP-BIC transforms into a high-Q quasi-BIC, which appears as a sharp Fano resonance (see Fig.~R1). In the absence of material losses, the transmission at the quasi-BIC frequency vanishes, and the single layer behaves as a perfect mirror. Consequently, in the bilayer configuration, additional Fabry-Pérot BICs emerge in the spectra, originating from these quasi-BIC resonances of the individual layers. However, we do not focus on this case in the present work.

\subsubsection{Bilayer structure}

Equations~\eqref{eq:refl} and \eqref{eq:tr} fully describe the scattering matrix of the single-layer structure. The scattering matrix of the double-layer system can be obtained either by applying the Redheffer star product~\cite{Redheffer1961} or by explicitly accounting for multiple reflections between the layers.

The real-valued poles of the scattering matrix correspond to the formation of FP-BICs. Alternatively, the eigenfrequency condition for the bilayer structure can be derived by requiring that, after a round trip between the layers, the amplitude of the mode remains the same and the accumulated phase changes by $2\pi m$, where $m$ is an integer~\cite{marcuse1991theory}. This condition can be written as
\begin{equation} \label{Eq:FP}
    r^2 e^{2i\omega L/c}=1.
\end{equation}

This equation can also be derived using the effective Hamiltonian approach, neglecting the near-field coupling between the layers (see Appendix~\ref{app:eff_H}). Substituting Eq.~\eqref{eq:refl} into Eq.~\eqref{Eq:FP} yields the characteristic equation for the eigenfrequencies:
\begin{equation}\label{eq:r_quantization}
   \frac{\pm\gamma_r}{i(\omega_0-\omega)+\gamma_r+\gamma_a}=e^{-i\omega L/c},
\end{equation}
Here ``$+$'' corresponds to an odd mode and ``$-$'' to an even one. This transcendent equation defines the resonances and their $Q$ factors in the bilayer structure as a function of $L$, $\gamma_a$, and $\gamma_r$. It can be solved numerically or using the perturbation theory. {We should note} 
that this equation does not describe the symmetry-protected BICs.

First, we set $\gamma_a = 0$ and find the distances $L$ ensuring the existence of FP-BICs. It can be shown that Eq.~\eqref{eq:r_quantization} {allows} 
real solutions $\omega = \omega_0$ for distances that satisfy the Fabry-Pérot quantization condition
\begin{equation}
    L_{m}=\frac {\pi m c}{\omega_0},\quad m=1,2,  \dots
    \label{eq:FP-quantization-L}
\end{equation}
where $m$ is the number of FP-BICs. Odd values of $m$ (i.e., $m=2k+1$) correspond to even FP-BICs, and even values of $m$ (i.e., $m=2k$){,} 
to odd FP-BICs. 
In contrast to a conventional Fabry-Pérot resonator, where the reflection coefficient of the mirrors is typically frequency-independent, the use of resonant mirrors introduces strong frequency dependence. In this case, only a single high-$Q$ mode is supported, which can be clearly identified in the spectrum due to its sharp and isolated resonance feature.

\subsection{Q factor, finesse, and mode volume}
If the interlayer distance deviates from $L_m$, defined by the Fabry-Pérot condition~\eqref{eq:FP-quantization-L}, by $\delta L$, it destroys the FP-BICs, transforming them into quasi FP-BICs with finite $Q$ factors. The total $Q$ factor becomes: 
\begin{equation}\label{eq:qtot}
1/Q_\text{tot}=1/Q^{\gamma_a}_\text{rad}+1/Q^{\gamma_a}_\text{abs}+1/Q^{L}_\text{rad}.   
\end{equation}
Here, $Q^{L}_\text{rad}$ corresponds to the radiation due to the deviation of the interlayer distance $\delta L$ from $L_m$;  $Q^{\gamma_a}_\text{abs}$, 
to the absorption induced by $\gamma_a$; {and} $Q^{\gamma_a}_\text{rad}$, 
to the radiative losses induced {by} $\gamma_a$. 

Assuming that $\gamma_a \ll \gamma_r$ and $\delta L / L_m \ll 1$, Eq.~\eqref{eq:r_quantization} can be analyzed to obtain analytical scaling laws for the different contributions to the total $Q$ factor. In this limit, one can separately identify: (i) radiative losses induced by absorption, (ii) direct absorption losses, and (iii) radiative leakage due to interlayer detuning: 
\begin{equation} \label{eq:Q_rad_a}
    Q^{\gamma_a}_\text{rad}=\pi\frac{\gamma_r^2}{\gamma_a^2}\frac{\omega_0 L_m^{\text{eff}}}{\pi c}=\mathcal{F}^{\gamma_a}_\text{rad}\frac{2L_m^{\text{eff}}}{\lambda_0},
\end{equation}
\begin{equation} \label{eq:Q_abs_a}
    Q_\text{abs}^{\gamma_a}=\frac{\pi}{2}\frac{\gamma_r}{\gamma_a}\frac{\omega_0 L_m^{\text{eff}}}{\pi c}=\mathcal{F}_\text{abs}^{\gamma_a}\frac{2L_m^{\text{eff}}}{\lambda_0},
\end{equation}
\begin{equation} \label{eq:Q_rad_L}
    Q^{L}_\text{rad}=\frac{\pi \gamma_r^2}{\omega_0^2 \overline{\delta L}^2}\frac{\omega_0 L_m^{\text{eff}}}{\pi c}=\mathcal{F}^{L}_\text{rad}\frac{2L_m^{\text{eff}}}{\lambda_0}.
\end{equation}
Here, $\mathcal{F}$ denotes the finesse of the Fabry-Pérot resonance {and} 
$2L_m^{\text{eff}}/\lambda_0$ {is} 
the one-dimensional effective mode volume normalized to half wavelength. The normalized shift is 
$\overline{\delta L} = \delta L / L_m^{\text{eff}}$. The one-dimensional effective mode volume is defined as

\begin{equation}
L_m^{\text{eff}}=L_m+c/\gamma_r.
\end{equation}
This expression differs from that of a conventional Fabry-Pérot resonator by the 
additional term, which arises due to the resonant nature of the mirrors. In the classical model of a Fabry-Pérot resonator with non-resonant mirrors, reflection is assumed to occur instantaneously, and the photon dwell time at the mirrors is negligible. In contrast, the mirrors considered in our system support resonances with a finite radiative lifetime $\tau_r = 1/\gamma_r$. As a result, photons are not reflected instantaneously, but 
temporarily trapped within the mirrors. This delayed reflection can be interpreted as an effective increase in the photon path length, equivalent to extending the physical distance between the mirrors by an additional length $c/\gamma_r$. Consequently, two limiting regimes can be identified: (i) a \emph{short-arm} Fabry--Pérot resonator, where $L_m \ll c/\gamma_r${, with} 
the photon energy 
predominantly confined within the resonant mirrors; and (ii) a \emph{long-arm} Fabry-Pérot resonator, where $L_m \gg c/\gamma_r${, with} 
the energy 
mainly stored in the cavity space between the mirrors.

\section{Effect of material losses}\label{sec:cmt}

\begin{figure}[t]
    \includegraphics[width=1\linewidth]{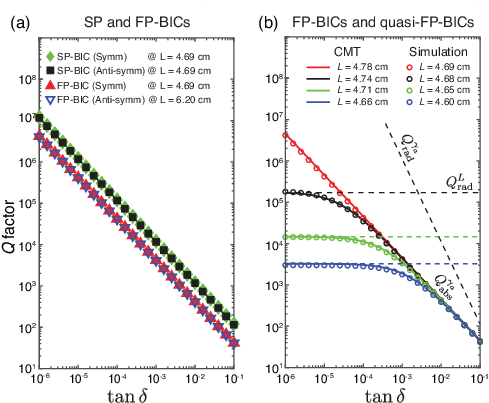}
    \caption{Dependence of $Q$ factors of different qBICs on loss tangent $\tan\delta$: (a) numerical results for the interlayer distance{s defined by the Fabry-Pérot quantization condition:} $L = 4.69$~cm for symmetry-protected qBICs and $B_{1u}$ FP-qBIC; 
    $L = 6.2$~cm for $B_{3g}$ FP-qBIC; (b) analytical and numerically simulated loss contributions to the $Q$ factor of the $B_{1u}$ FP-qBIC, including asymptotics given by Eqs.~\eqref{eq:Q_rad_a}, \eqref{eq:Q_abs_a}, and~\eqref{eq:Q_rad_L}, {for interlayer distance detuned from the Fabry-Pérot resonance condition.}}
    \label{fig:q-tand}
\end{figure}

Let us analyze the impact of material losses on the symmetry-protected BICs ($A_u$, $B_{2g}$). Since the introduction of material absorption does not break the C$_2$ symmetry of the system, these BICs persist but acquire a finite $Q$ factor. In comparison to quasi-FP-BIC, which $Q$ factor is affected by contributions described in Eq.~\eqref{eq:qtot}, the conservation of C$_2$  symmetry prohibits the radiative losses induced $\gamma_a$. Moreover, quasi SP-BIC are robust to deviations in the interlayer distance, and the only contribution to the total $Q$ factor is absorptive. Figure~\ref{fig:q-tand}(a) shows the dependence of the total quality factor{s} $Q_\text{tot}$ {of the} 
symmetry-protected and FP-BICs {on the loss tangent for the interlayer distances defined by the Fabry-Pérot quantization condition}. For both types of BICs, the absorptive contribution to the total $Q$ factor follows the relation

\begin{equation}
    Q_\text{abs}^{-1} \approx  \Gamma \tan\delta,
\end{equation} \vspace*{-12pt}
\begin{equation}
    \Gamma_{m} = \frac{\int_{\text{rod}} \varepsilon'(\mathbf{r}) |\mathbf{E}(\mathbf{r})|^2 \, d\mathbf{r}}{\int_{\text{space}} \varepsilon'(\mathbf{r}) |\mathbf{E}(\mathbf{r})|^2 \, d\mathbf{r}},
\end{equation}
where $\Gamma_{m}$ {is} 
the optical confinement factor of the mode~\cite{coldren2012diode}. In our case, due to the low refractive index and sparse geometry of the structure, $\Gamma_{m}$ is approximately 0.1 for all {the considered} BICs. 
In contrast, for structures composed of high-index materials such as ceramics or water, the confinement factor approaches unity ($\Gamma_{m} \approx 1$), which significantly limits the achievable $Q$ factors~\cite{odit2021observation,bogdanov2019bound}.

Figure~\ref{fig:q-tand}(b) shows the dependence of the {$Q$} factor of quasi-FP-BICs on loss tangent $\tan\delta$ when the interlayer distance is detuned from the optimal value $L_m$, defined by the quantization condition in Eq.~\eqref{eq:FP-quantization-L}. The red solid line in Fig.~\ref{fig:q-tand}(b) corresponds to the case when the interlayer distance matches the quantization condition perfectly. The solid lines  are the results obtained from CMT, while the markers indicate data from full-wave numerical simulations (COMSOL Multiphysics). The analytical and numerical results are in good agreement. The dashed lines are the contributions of different loss mechanisms to the total $Q$ factor given by Eqs.~\eqref{eq:Q_rad_a}, \eqref{eq:Q_abs_a}, and \eqref{eq:Q_rad_L}. For small values of the loss tangent $\tan\delta$, the radiative losses dominate, whereas for larger values $\tan\delta$, absorption becomes the primary loss mechanism for quasi-FP-BICs. As seen from Eqs.~\eqref{eq:Q_rad_a}–\eqref{eq:Q_rad_L}, the different loss-induced contributions to the FP-qBIC quality factor exhibit distinct scaling behaviors with respect to the material loss $\gamma_a$. The absorption-induced radiative term obeys $Q_{\text{rad}}^{(\gamma_a)} \propto 1/\gamma_a^{2}$, whereas the purely absorptive term scales as $Q_{\text{abs}}^{(\gamma_a)} \propto 1/\gamma_a$ (with $\gamma_a \propto \tan\delta$). Consequently, at small $\tan\delta$, the radiative channel dominates the total $Q$ [Eq.~\eqref{eq:qtot}], while at larger $\tan\delta$, the absorptive channel prevails and the total quality factor decreases approximately as $Q_\text{tot}\sim Q_{\text{abs}}^{(\gamma_a)} \propto 1/\tan\delta$. This crossover explains the monotonic reduction of $Q$ with increasing loss tangent observed in Fig.~\ref{fig:q-tand}(b), except for the exactly tuned FP–BIC case, where $\delta L=0$ and the no-absorption limit yields $Q_\text{tot} \to \infty$. The critical coupling regime corresponds to balanced radiative and non-radiative losses, leading to maximum field enhancement. The balance condition can be derived analytically from Eqs.~\eqref{eq:Q_abs_a} and~\eqref{eq:Q_rad_L}:
\begin{equation}\label{eq:crit_coupling}
    2\gamma_r \gamma_a = (\omega_0 \, \overline{\delta L})^2.
\end{equation}

As discussed in Sec.~\ref {sec:single-layer-structure}, FP-BICs are inherently non-robust {against} 
material losses. 
Absorption makes the resonant mirrors partially transparent, preventing perfect confinement of the mode and thereby transforming {an} 
FP-BIC into a quasi-FP-BIC [see Eq.~\eqref{eq:part_transparencey}]. This instability arises because an FP-BIC, as an accidental BIC, is inherently sensitive to time-reversal symmetry breaking, and, therefore, it is generally destroyed by material losses~\cite{zhen2014topological,hsu2013observation}. However, the radiative losses induced by material absorption appear only as a second-order correction,
i.e. the associated radiative $Q$ factor scales as $Q_\text{rad}^{\gamma_a} \propto 1/\gamma_a^2$ [see Eq.~\eqref{eq:Q_rad_a}]. Consequently, these induced radiative losses remain almost negligible in a broad range of $\tan\delta$ [see the corresponding dashed line in Fig.~\ref{fig:q-tand}(b)].

\section{Effect of interlayer distance}
Let us analyze how the interlayer distance $L$ between the resonant mirrors affects the radiative and absorption losses of symmetry-protected and Fabry-Pérot BICs in the bilayer structure. As a representative case, we fix the loss tangent at $\tan\delta = 0.01$.

\begin{figure}[tb]
    \includegraphics[width=1\linewidth]{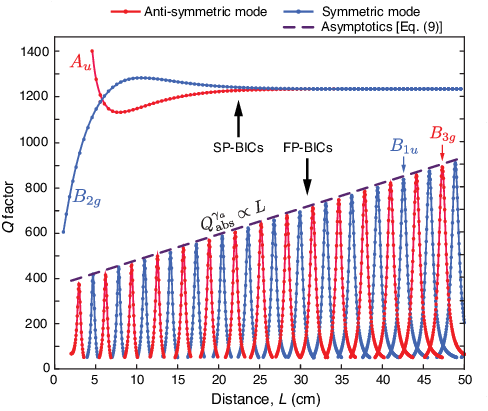}
    \caption{Dependence of $Q$ factors of different qBICs on the interlayer distance $L$ of dielectric rods with fixed losses $\tan\delta=0.01$ and $Q_\text{abs}^{\gamma_a}$ asymptotics given by Eq.~\eqref{eq:Q_abs_a}.}
    \label{fig:Qh_withlosses}
\end{figure}

At small interlayer distances, symmetry-protected BICs localized in each layer interact via near-field, which modifies the mode profiles and their overlap with the absorbing material. This interaction leads to the $Q$ factor oscillating with 
$L$, as shown in Fig.~\ref{fig:Qh_withlosses}. In particular, for small values of $L$, the symmetric BIC ($B_{2g}$) {field is strongly confined}  inside the dielectric rods, resulting in increased absorption losses. In contrast, the field of anti-symmetric BIC ($A_u$) {weakly} overlap{s} with the rods {i.e.~there are} lower losses and a lower effective mode index, {leading to a} blue spectral shift. When $L \approx 4.5$~cm, this frequency enters the domain where $\pm 1$ diffraction 
{channels open}, thereby destroying the BIC. As the distance increases, the near-field interaction becomes negligible for $L \gg p$, where $p$ is for period [see Fig.~\ref{fig:Qh}(a)], and the symmetry-protected BICs in each layer effectively decouple. Consequently, the $Q$ factors of both symmetric and anti-symmetric symmetry-protected BIC become independent of $L$ (see Fig.~\ref{fig:Qh_withlosses}).

Fabry-Pérot BICs exhibit {a} fundamentally different behavior. Their electromagnetic energy is confined within the resonant mirrors and in the cavity region between them. As the interlayer distance $L$ increases, the energy stored between the mirrors grows linearly, while the material losses -- occurring {only in} 
the mirrors -- remain constant. {As a result,} 
the $Q$ factor of FP-BICs {linearly increases} with 
$L$. Consequently, the $Q$ factor of FP-BICs can become arbitrarily large, even in systems with material absorption. The analytical model described by Eq.~\eqref{eq:Q_abs_a} shows excellent agreement with the numerical results obtained using COMSOL, as illustrated by the {dashed} violet 
line in Fig.~\ref{fig:Qh_withlosses}.

We should also note that variations in the inter-mirror distance spectrally shift the Fabry-Pérot modes. For small normalized shifts, $\overline{\delta L} \ll 1$, the resonance
frequency can be approximated as
\begin{equation}
\omega = \omega_0(1 - \overline{\delta L} + \overline{\delta L}^2).
\end{equation}
Notably, corrections to the imaginary part of $\omega$ arise only in the second-order perturbation theory with respect to $\overline{\delta L}$.

\begin{figure}[bt]
    \includegraphics[width=1\linewidth]{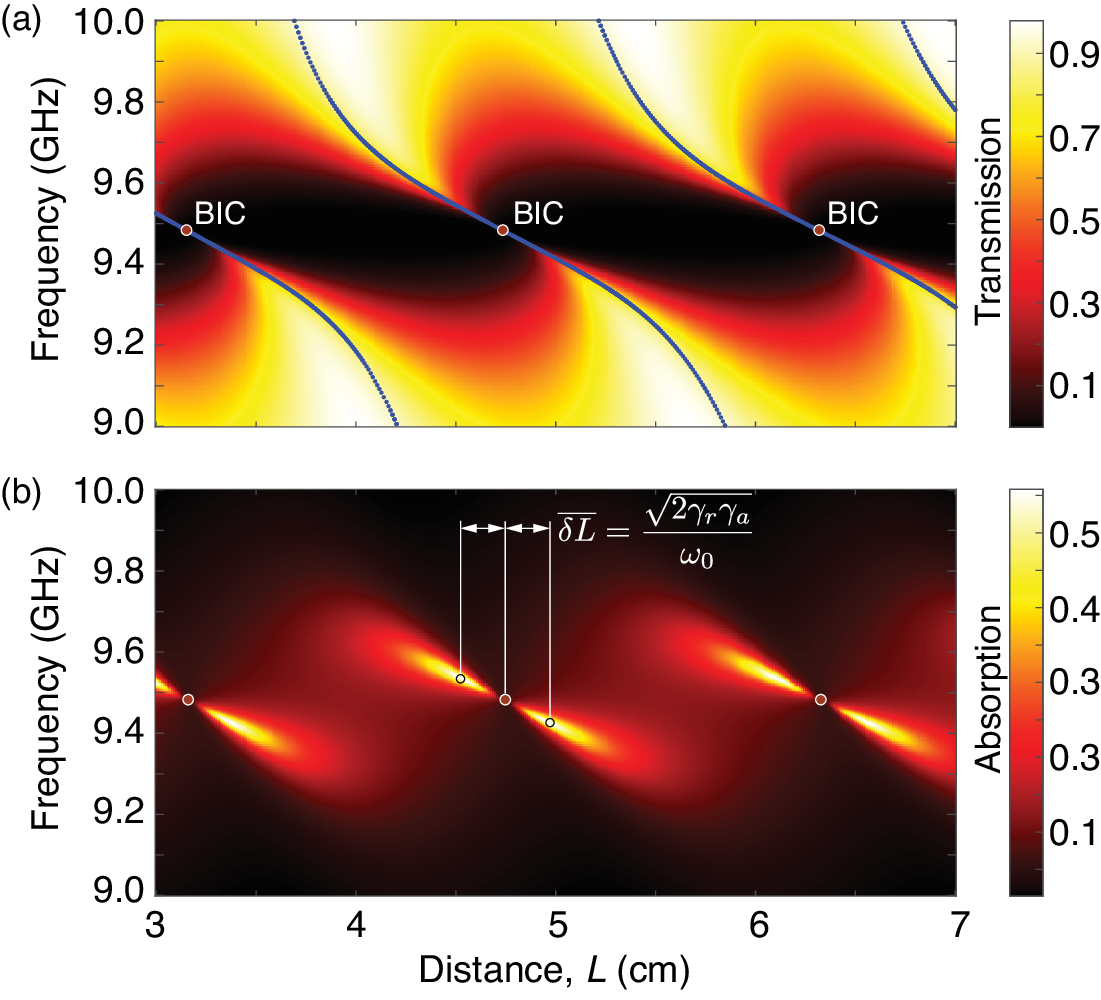}
    \caption{Impact of relative shift of the layers: 
    (a) Transmission and (b) absorption map of the bilayer structure as a function of {interlayer} distance 
    $L$ calculated with CMT. The absorption loss {is} $\gamma_a$=0.01. 
    {Solid} blue 
    lines in panel (a) show 
    the spectral position of the Fabry-Pérot resonances. The {absorption} maxima 
    in panel (b) correspond {to} the critical coupling regime defined by Eq.~\eqref{eq:crit_coupling}.}
    \label{fig:tr_and_abs}
\end{figure}

Figure~\ref{fig:tr_and_abs} presents the transmission [panel~(a)] and absorption [panel~(b)] spectra of the bilayer structure as functions of the interlayer distance $L$. The solid blue 
lines in panel~(a) indicate the spectral positions of the Fabry-Pérot resonances. As shown in panel~(b), the absorption exhibits pronounced maxima at specific values of $\pm\overline{\delta L}$, corresponding to the critical coupling condition defined by Eq.~\eqref{eq:crit_coupling}. Although critical coupling enhances absorption, the latter does not reach unity due to radiation in both upward and downward directions. At the BIC positions, absorption remains nonzero but extremely small, because the effective radiative losses introduced by absorption scale as $1/\gamma_a^2$ [see Eq.~\eqref{eq:Q_rad_a}].

\section{Effect of layer displacement} 

\begin{figure}[t!]
    \includegraphics[width=1\linewidth]{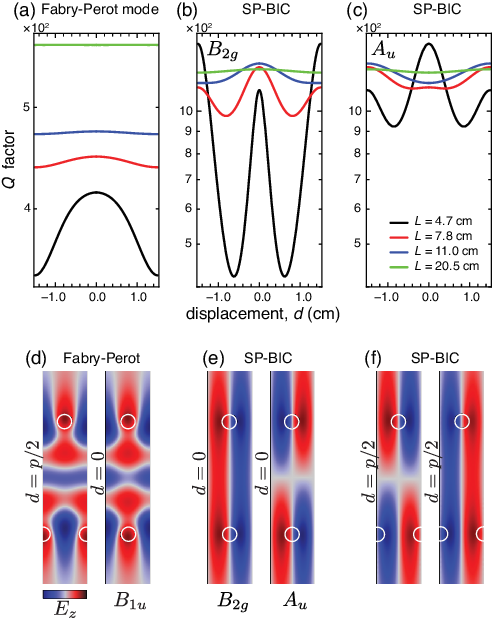}
    \caption{Impact of relative displacement of the layers: 
(a) Dependence of the $Q$ factor of the $B_{1u}$ FP-BIC on the displacement $d$ for $L=4.7$~cm (solid black line), $7.8$~cm (solid red line), $11.0$~cm (solid blue line), and $20.5$~cm (solid green
    line); (b) Dependence of the $Q$ factor of the symmetry-protected $B_{2g}$ modes on the displacement $d$; (c) Dependence of the $Q$ factor of the symmetry-protected $B_{1g}$ modes on the displacement $d$. (d) Field distribution of the $B_{1u}$ mode at $d=0$ (FP-BIC) and its symmetry-broken counterpart at $d=p/2=1.5\,\text{cm}$ for $L=7.8\,\text{cm}$; (b) Field distribution of the symmetry-protected BICs $B_{2g}$ and  $A_{u}$ at $d=0$ and (f) their counterparts at $d=p/2=1.5\,\text{cm}$ for $L=7.8\,\text{cm}$. The loss tangent is $\tan\delta = 0.01$ in all panels.}
    \label{fig:Qshift}
\end{figure}

In a bilayer system, the $Q$ factor of BICs depends on the relative alignment of the layers, described by the displacement parameter $d \in [-p/2, p/2]$, which acts as a perturbation. Here $p$ is the period. This perturbation preserves inversion symmetry and does not introduce a bianisotropic response~\cite{poleva2023multipolar}. To illustrate the effect of displacement for more realistic structures, we fix the loss tangent to $\tan\delta = 0.01$. 

Figure~\ref{fig:Qshift}(a) shows the dependence of the $Q$ factor of the symmetric FP-BIC $B_{1u}$ [Fig.~\ref{fig:Qshift}(d)] for different interlayer distances $L$. For all values of $L$, the $Q$ factor decreases monotonically with $d$. However, for large $L$, the $Q$ factor becomes almost insensitive to displacement. This behavior arises because the total field between the layers for FP-BICs contains both near- and far-field components. As the interlayer distance $L$ increases, the near-field contribution decays and the layers interact predominantly through plane-wave (far-field) components. In this regime, the system becomes much more robust, i.e., almost insensitive to lateral displacement along the periodicity axis. For the SP-BICs $B_{2g}$ and $A_u$, the trend is similar [Figs.~\ref{fig:Qshift}(b) and ~\ref{fig:Qshift}(c)] -- the sensitivity of the $Q$ factor to $d$ decreases with increasing $L$. This is because SP-BICs in each layer interact only through the near-field, and at sufficiently large $L$ the layers effectively decouple and the SP-BICs behave independently.

\begin{figure}[t!]
    \includegraphics[width=1\linewidth]{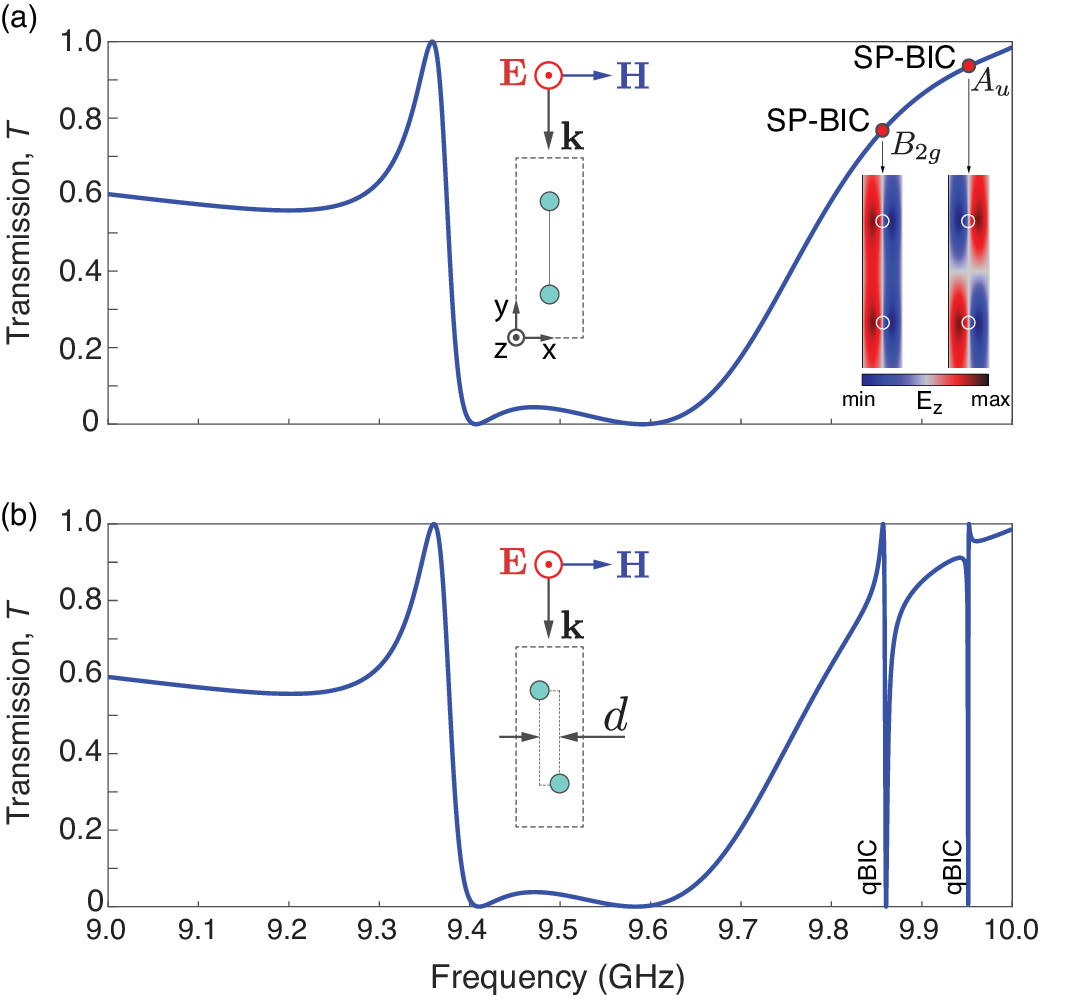}
    \caption{Transmission spectra of the bilayer dielectric rod array in the lossless limit ($\tan\delta = 0$). 
(a) Aligned configuration ($d = 0~\text{cm}$), where two symmetry-protected BICs of $A_u$ and $B_{2g}$ symmetry appear at normal incidence.  (b) Laterally displaced configuration with $d = 0.1~\text{cm}$, in which the $C_2$ symmetry is broken and the SP-BICs transform into high-$Q$ quasi-BICs, visible as sharp Fano-type resonances.}
    \label{fig:displacement}
\end{figure}

For both symmetry-protected BICs, two maxima of the $Q$ factor are observed at $d = 0$ and $d = p/2$ [Figs.~\ref{fig:Qshift}(b) and \ref{fig:Qshift}(c)]. The $Q$ factor of the $B_{2g}$ mode at $d = 0$ coincides with that of the $A_{u}$ mode at $d = p/2$, and vice versa. The displacement generally breaks the $yz$-mirror symmetry and destroys the SP-BICs. However, the case $d = p/2$ deserves special attention. At $d=p/2$, the $yz$-mirror symmetry is restored, but the point-group symmetry is reduced from $D_{2h}$ (at $d=0$) to $D_1$ (at $d=p/2$). Nevertheless, two symmetry-protected BICs still exist at $d=p/2$~\cite{Ndangali2010electromagnetic}, and their field profiles remain similar to those at $d=0$ [see Figs.~\ref{fig:Qshift}(e) and \ref{fig:Qshift}(f)]. This behavior can be understood in terms of frieze groups~\cite{tilleyCrystalsCrystalStructures2006,martinez2024pumping}. For $d=0$ and $d=p/2$, the spatial symmetry of the bilayer structure corresponds to the frieze groups $p2mm$ and $p2mg$, respectively. These two frieze groups are isomorphic, and both admit the existence of two types symmetry-protected BICs.

Figure~\ref{fig:displacement} shows the transmission spectra of the bilayer dielectric rod array for (a) the aligned configuration and (b) the laterally displaced configuration. In panel (b), a lateral shift of $d = 0.1~\text{cm}$ is introduced between the layers, and the simulations are performed in the lossless limit ($\tan\delta = 0$). In the aligned structure, two symmetry-protected BICs of $A_u$ and $B_{2g}$ symmetry are not observed at normal incidence. When the displacement is applied, the $C_2$ symmetry is broken and the SP-BICs are destroyed, giving rise to high-$Q$ quasi-BIC resonances that appear as sharp Fano features in the spectrum.

\section{Effect of Oblique Incidence} \label{app:oblique}

\begin{figure}[t!]
    \includegraphics[width=1\linewidth]{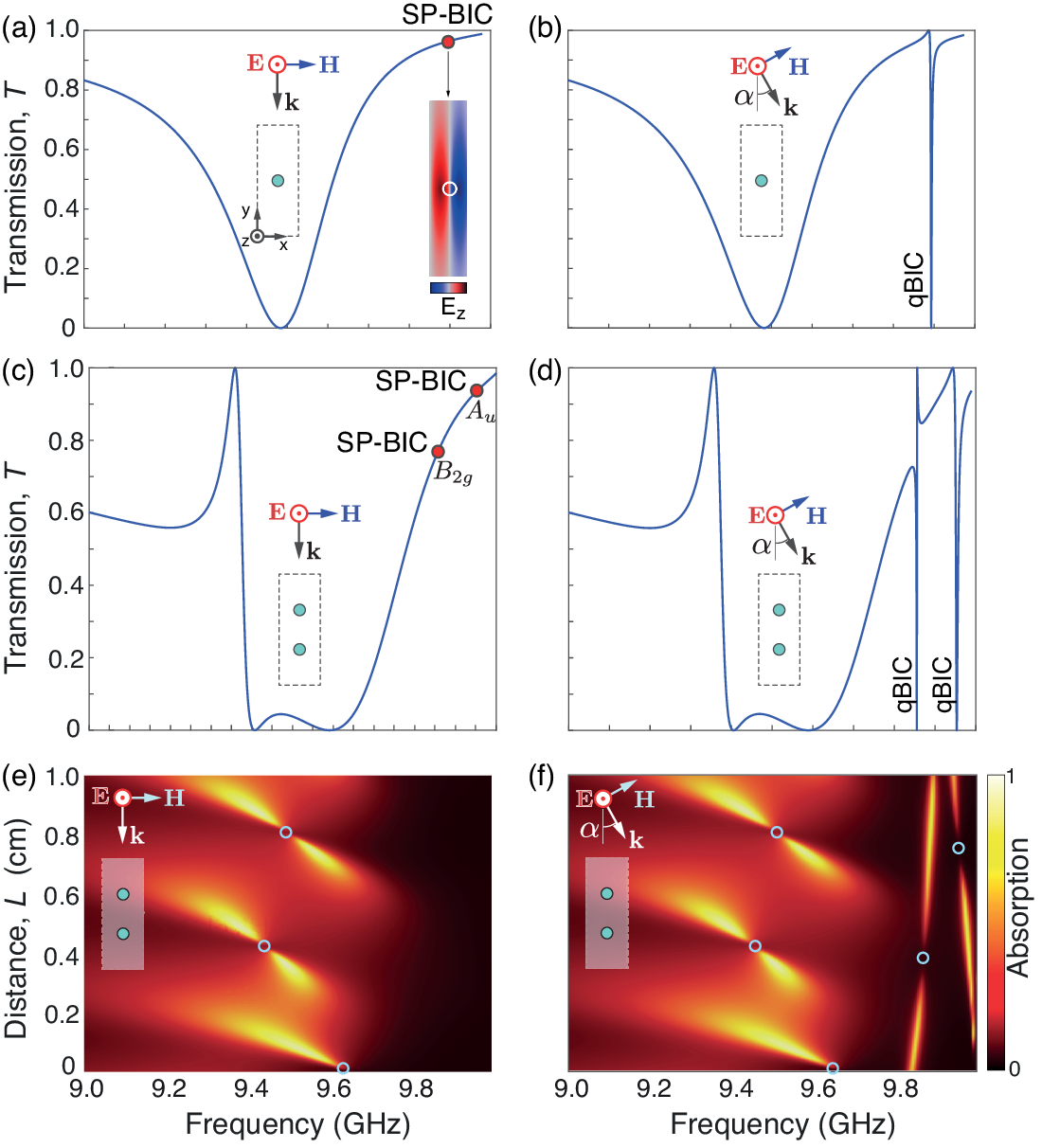}
    \caption{Transmission spectra of single-layer and bilayer dielectric rod arrays under normal and oblique incidence. 
Panels (a)--(d) correspond to the lossless case ($\tan\delta = 0$). 
(a) Single layer at normal incidence: the SP-BIC remains uncoupled and does not appear in the spectrum. 
(b) Single layer at oblique incidence ($\alpha = 0.2^\circ$): the SP-BIC becomes a high-$Q$ quasi-BIC and appears as a sharp Fano resonance. 
(c) Bilayer structure supporting two SP-BICs ($A_u$ and $B_{2g}$) at normal incidence. 
(d) Under oblique incidence ($\alpha = 0.2^\circ$), both SP-BICs convert into quasi-BICs. 
Panels (e)--(f) show absorption maps for $\tan\delta = 10^{-3}$: 
(e) normal incidence and (f) oblique incidence ($\alpha = 0.2^\circ$).}
    \label{fig:oblique}
\end{figure}

In this section, we analyze the behavior of symmetry-protected and Fabry-Pérot BICs under oblique incidence and compare it with the normal-incidence case, as summarized in Fig.~\ref{fig:oblique}. A single-layer rod array supports a symmetry-protected BIC at normal incidence, however, this state does not manifest itself in the transmission spectrum because it is completely decoupled from the radiation continuum [see Fig.~\ref{fig:oblique}(a)]. When a small oblique angle is introduced, the symmetry protection is lifted ($k_x\neq0$) and the BIC transforms into a high-$Q$ quasi-BIC. This quasi-BIC appears as a sharp Fano resonance, as demonstrated in Fig.~\ref{fig:oblique}(b). 

In the bilayer configuration, two symmetry-protected BICs arise at normal incidence, corresponding to the symmetric ($A_u$) and antisymmetric ($B_{2g}$) combinations of the single-layer BICs [Fig.~\ref{fig:oblique}(c)]. As in the single-layer case, these SP-BICs remain invisible in the transmission spectrum under normal incidence. Under oblique incidence, however, both modes convert into high-$Q$ quasi-BICs and manifest as pronounced Fano resonances [Fig.~\ref{fig:oblique}(d)].

In the lossless limit, the transmission near the quasi-BIC in the single-layer array vanishes, and the layer behaves as a perfect mirror. As a result, a pair of such layers can form Fabry-Pérot BICs at interlayer distances $L$ that satisfy the Fabry-Pérot quantization condition. In contrast to the normal-incidence case, additional FP-BICs appear at oblique incidence. These FP-BICs originate directly from the quasi-BICs of the individual layers. This behavior is illustrated in the absorption maps shown in Figs.~\ref{fig:oblique}(e) and \ref{fig:oblique}(f), where the absorption is plotted versus frequency and interlayer distance $L$ for normal and oblique incidence ($\alpha=0.2\degree$). Under oblique incidence [Fig.~\ref{fig:oblique}(f)], additional Fabry-Pérot branches appear due to the quasi-BICs. These features form clear Fabry-Pérot-type trajectories in the absorption map, demonstrating that oblique incidence enables an additional mechanism for generating FP-BICs in multilayer structures.

\section{Conclusions} 

In conclusion, we have addressed how material and geometric perturbations affect the $Q$ factor of symmetry-protected and Fabry-Pérot quasi-BICs in a bilayer periodic array of infinitely long dielectric rods. We have determined that introducing material losses into structural elements significantly impacts the $Q$ factor for both types of BICs. The study distinguishes and compares the contributions of different loss mechanisms to the $Q$ factor of FP-BICs. In particular, absorption reduces the $Q$ factor of the resonator more significantly than other loss types. For symmetry-protected BICs with a fixed loss tangent, the $Q$ factor remains unaffected by changes in the interlayer distance. In contrast, an increased interlayer distance increases the $Q$ factor for FP-BICs.  Furthermore, lateral displacement of one layer leads to significant decreases in the $Q$ factors of FP-BICs and SP-BICs. Yet, displacements by half the period restore C$_2$ symmetry of the system and allow for SP-BICs with high $Q$ factor. BICs also become less sensitive to C\textsubscript{2}-breaking perturbations as the interlayer distance increases. In general, our results underscore the intricate interplay between material losses, geometry, and the resulting $Q$ factors for different BIC types in the analyzed system.

\begin{acknowledgments} \label{sec:acknowledgments}
We acknowledge the support of the Russian Science Foundation. The numerical studies were supported by the Russian Science Foundation (Project 25-12-00261) and the analytical studies were supported by the Russian Science Foundation (Project 25-42-10025). A.B. acknowledges support from the National Natural Science Foundation of China (Project W2532010). The authors thank Lydia Pogorelskaya for her proofreading of the English manuscript. 
\end{acknowledgments}

\bibliography{bic}

\appendix
\section{Bloch{'s} theorem}\label{app:Bloch}

As mentioned above, symmetry is a mathematical condition for the emergence of BIC. According to Bloch's theorem, the electric field of the eigenmodes in a periodic structure can be expressed as:
\begin{equation}
    \vec{E}(x,y,z)=\vec{U}_{l,k_x}(x,y)e^{ik_xx+ik_zz},
\end{equation}
where $\vec{U}_{l,k_x}(x,y)$ is a periodic function with the same periodicity as the lattice, $k_x$ is the wave vector associated with the propagation of the mode, and $l$ is the index of the band, which we omit here for simplicity. This representation illustrates that the electric field is composed of a plane wave component modulated by a periodic function, indicative of the periodicity of the structure.

In such a system, BICs appear only if $k_z=0$ or $k_x=0$. We further limit our analysis to the case of TE-polarized modes [$\vec{E}=(0,0,E_z)$] with $k_z=0$. For TE modes, $k_z=0$, we can write $E_{z}(x,y) = U_{k_x}(x,y) e^{i k_x x}$. The periodic Bloch amplitude can be expanded into the Fourier series as follows:
\begin{equation}
     U_{k_x}(x,y) = \sum_n C_{n,k_x}(y) e^{i\frac{2\pi n}{a}x},
\end{equation}
where $n$ is the index of the diffraction channel. At frequencies above the light line ($\omega/c>|k_x|$), the mode leaks from the structure to the radiation continuum via the open diffraction channels. BICs arise when leakage into all the open diffraction channels is prohibited, which means that the complex Fourier coefficients $C_{n,k_x}(y)$ - the amplitudes of the outgoing waves — are zero. In the subwavelength regime, 
$\lambda > a$, only the zeroth diffraction channel remains open. Consequently, the amplitude of the outgoing leaky wave is determined by the zeroth Fourier coefficient $C_{0,k_x}(y)$, which is defined as the field component averaged over the period: $C_{0,k_x}(y) = \langle U_{0,k_x}(x,y)\rangle_x$.

For structures exhibiting time-reversal and $\pi$-rotational symmetries, referred to as $TC_{2}^{y}$, the coefficient $C_{0,k_x}(y)$ becomes purely real for BICs \cite{hsu2013observation,PhysRevLett.118.267401,zhen2014topological}. At the $\Gamma$-point (the center of the Brillouin zone), $U_{k_x}(x,y)$ is either an odd or even function, as the photonic structure is $C_2^y$-invariant~\cite{Sakoda_2001,Ivchenko_Pikus_Skrebtsov_1995}. 

For an odd function $U_{k_x}(x,y)$, the zero-order Fourier coefficient vanishes, leading to a symmetry-protected BIC, with coupling to the radiation continuum 
eliminated due to the 
point symmetry {of the system}. For an even mode, the spatial average $\langle U_{0,k_x}(x,y)\rangle_x$ may vanish not only because of symmetry, but also {for} 
specific values of geometric and material structural parameters, resulting in the so-called accidental BIC or tunable BIC. In the $\Gamma$-point case, these resonances are called Fabry-Pérot BICs.

\section{Symmetry of modes}\label{app:symmetry}

Since the bilayer resonator possesses
$D_{2h}$ symmetry, we analyzed the modes of the resonator and classified them according to their point groups using the characteristic table \cite{gelessus1995multipoles}. The symmetric symmetry-protected mode, denoted as $B_{2g}$, retains $C^{y}_{2}$, $\sigma_{v}$, and inversion symmetries. Further, the anti-symmetric symmetry-protected mode, denoted as $A_{u}$, is characterized by $C_{2}$ symmetry for all orientations. In contrast, the symmetric FP-mode demonstrates $C^{y}_{2}$, $\sigma_{h}$, and $\sigma_{d}$ symmetries, which means it has $B_{1u}$ symmetry. The anti-symmetric FP-mode possesses $C^{x}_{2}$ symmetry and inversion through a center of symmetry and is denoted as $B_{3g}$. The 
classification {of the modes according to their symmetry} is presented in Table~\ref{tab:symmetry}. 

\begin{table}[t]
\renewcommand{\arraystretch}{1.3}
\caption{$D_{2h}$ point group representations.}
\begin{center}
\begin{tabular}{|c|c|}
\hline
Mode & Representation \\
\hline
Symmetry-protected symmetric & $B_{2g}$ \\
\hline
Symmetry-protected anti-symmetric & $A_{u}$ \\
\hline
Fabry-P\'erot symmetric & $B_{1u}$ \\
\hline
Fabry-P\'erot anti-symmetric & $B_{3g}$ \\
\hline
\end{tabular}
\end{center}
\label{tab:symmetry}
\end{table}

\section{Effective Hamiltonian approach}\label{app:eff_H}
Here, we analyze a
simple Hamiltonian describing a two-resonance photonic system~\cite{Sadreev2021interference,Shabanov2009resonant,CanosValero2025exceptional}:
\begin{equation}
    \hat{H}=\begin{pmatrix}
        \omega_1 & \kappa\\
        \kappa & \omega_2
    \end{pmatrix} - i \begin{pmatrix}
        \gamma_1 & e^{i\phi}\sqrt{\gamma_1\gamma_2} \\
        e^{i\phi}\sqrt{\gamma_1\gamma_2} & \gamma_2
    \end{pmatrix},
\end{equation}
where $\omega_{1,2}$ are the resonance frequencies of the coupled modes, $\kappa$ is the near-field internal coupling, $\gamma_{1,2}$ are the radiative losses of the modes, and $\phi$ is the phase delay.

If a state $(a, b)^T$ is a BIC, then its eigenfrequency must be purely real. Thus, we can write
\begin{equation}
    \begin{pmatrix}
        \gamma_1 & e^{i\phi}\sqrt{\gamma_1\gamma_2} \\
        e^{i\phi}\sqrt{\gamma_1\gamma_2} & \gamma_2
    \end{pmatrix}\begin{pmatrix}
        a\\b
    \end{pmatrix}=\begin{pmatrix}
        0\\0
    \end{pmatrix},
\end{equation}
which leads to the condition
\begin{equation}
    e^{2i\phi}=1,\quad \phi=\pi m,\; m\in\mathbb Z,
\end{equation}
and the eigenvector can be expressed as
\begin{equation}
    \begin{pmatrix}
        a\\b
    \end{pmatrix} = \begin{pmatrix}
        \sqrt{\gamma_2} e^{i\phi} \\ -\sqrt{\gamma_1}
    \end{pmatrix}.
\end{equation}

Finally, eliminating
the unknown eigenfrequency $\omega$ from the equation
\begin{equation}
    \begin{pmatrix}
        \omega_1 & \kappa\\
        \kappa & \omega_2
    \end{pmatrix}\begin{pmatrix}
        \sqrt{\gamma_2} e^{i\phi} \\ -\sqrt{\gamma_1}
    \end{pmatrix} = \omega \begin{pmatrix}
        \sqrt{\gamma_2} e^{i\phi} \\ -\sqrt{\gamma_1}
    \end{pmatrix}
\end{equation}
yields another necessary condition for the system parameters
\begin{equation}
    \kappa(\gamma_1-\gamma_2)=\exp\left(i\phi\right) \sqrt{\gamma_1\gamma_2}(\omega_1-\omega_2).
\end{equation}

In the special case when $\omega_1=\omega_2=\omega_0$, $\gamma_1=\gamma_2=\gamma_r$, and $\kappa=0$, diagonalizing $\hat H$ yields the following restriction for the eigenfrequencies:
\begin{equation}
    (\omega-\omega_0+i\gamma_r)^2=\left(ie^{i\phi}\gamma_r\right)^2,
\end{equation}
resulting in
\begin{equation}
    e^{-i\phi}=\pm\frac{\gamma_r}{i(\omega-\omega_0)+\gamma_r}.
\end{equation}

\section{CMT and S-matrix}\label{app:cmt}
Notably, the {term} FP-BIC 
{stems from the formation} mechanism of {these BICs,} 
similar to {that of} resonances in a Fabry-Pérot resonator, in which the field is trapped between two perfect mirrors. Thus, to describe FP BICs
in the structure of two layers of dielectric rods, we need
to calculate the transmission through
each layer and determine its
minima, which will define
the ``mirror'' states of a layer. Below, we present the coupled mode theory (CMT) for this problem.

\subsection{CMT for a single layer}

The time-harmonic steady-state coupled-mode equation for a resonant mirror can be written as
\begin{equation}\label{eq:app-dadt}
-i\omega a = -i\omega_0 a - \gamma a +
\underbrace{
\begin{bmatrix}
 \sqrt{\gamma_r} e^{i\phi} &  \sqrt{\gamma_r} e^{i\phi}
\end{bmatrix}}_{\mathbf{K}^T }
\begin{bmatrix}
s_{1+} \\
s_{2+}
\end{bmatrix},
\end{equation}
where $\omega_0$ is the resonant frequency; $\gamma=\gamma_r+\gamma_a$, $\gamma_r$ is the radiation loss, and $\gamma_a$ is the absorption loss. The outgoing waves are given by
\begin{equation}\label{eq:app-outgoing}
\begin{bmatrix}
s_{1-} \\
s_{2-}
\end{bmatrix}
=
\underbrace{\begin{bmatrix}
0 & -e^{2i\phi} \\
-e^{2i\phi} & 0
\end{bmatrix}}_{{\hat C}}
\begin{bmatrix}
s_{1+} \\
s_{2+}
\end{bmatrix}
+
\underbrace{\begin{bmatrix}
\sqrt{\gamma_r} e^{i\phi} \\
\sqrt{\gamma_r} e^{i\phi}
\end{bmatrix}}_{\mathbf{D}}
a,
\end{equation}
where $\mathbf{K}$ and $\mathbf{D}$ are the coupling and decoupling vectors, and $\hat C$ is the 
direct scattering {matrix}. Without loss of generality, we can assume that 
$e^{2i\phi}=-1$. 
Then the transmission and reflection amplitude coefficients can be written as
\begin{equation}\label{eq:eq28-sup}
    r=\frac{-\gamma_r}{i(\omega_0-\omega)+\gamma_r+\gamma_a},
\end{equation}
\begin{equation}\label{eq:eq28-sup-1}
    t=\frac{i(\omega_0-\omega)+\gamma_a}{i(\omega_0-\omega)+\gamma_r+\gamma_a}.
\end{equation}
Here, $s_{i+}$ and $s_{i-}$ denote
the incident wave and the reflected wave at the $i$th port, respectively, $\omega_0$ is the resonant frequency, $\gamma_r$ is the radiation loss, and $\gamma_a$ is the absorption loss.

\subsection{CMT for two layers}

Similarly to the single-layer case, let us {consider} 
two identical mirrors and two modes $a^{1,2}$:
\begin{equation}
    -i\omega a^j = -i\omega_0 a^j - \gamma a^j + \vec K^T \begin{bmatrix}
    s^j_{1+} \\
    s^j_{2+}
    \end{bmatrix}
\end{equation}
and, accordingly, the outgoing waves are
\begin{equation}
    \begin{bmatrix}
        s_{1-}^j \\ s_{2-}^j
    \end{bmatrix} = \hat C \begin{bmatrix}
        s_{1+}^j \\ s_{2+}^j
    \end{bmatrix} + \vec D a^j,
\end{equation}
where all {the} definitions are the same as in~\eqref{eq:app-dadt} and~\eqref{eq:app-outgoing}, and $j=1$ or $2$.

Additionally, 
the waves traveling between the mirrors {are related as}:
\begin{equation}\label{eq:app-connect}
    \begin{bmatrix}
        s_{2-}^1 \\ s_{1-}^2
    \end{bmatrix} = e^{i\phi} \begin{bmatrix}
        s_{1+}^2 \\ s_{2+}^1
    \end{bmatrix}.
\end{equation}

The condition~\eqref{eq:app-connect} will be satisfied if and only if it is true that
\begin{equation}
    \gamma_r e^{i\phi}=i(\omega_0-\omega)+\gamma_r+\gamma_a,
\end{equation}
or, since $\phi=\omega L/c$,
\begin{equation}
    \gamma_r e^{i\omega L/c}=i(\omega_0-\omega)+\gamma_r+\gamma_a.
\end{equation}

\subsection{S-matrix for single-layer and bilayer structures} 
The scattering matrix $\hat{S}$ of a single-layer structure can be expressed in terms of the amplitude reflection ($r$) and transmission ($t$) coefficients, as defined in Eqs.~\eqref{eq:eq28-sup} and \eqref{eq:eq28-sup-1}, and takes the form~\cite{haus1984waves}:
\begin{equation}
\hat S=
\begin{bmatrix}
r & -t \\
-t & r \\
\end{bmatrix}.
\end{equation}

The full Fabry-Pérot resonator $\hat S_\text{FP}$ can be {described by} 
the Redheffer star product of three components~\cite{Redheffer1961} 
\begin{equation}
\hat S_\text{FP}=\hat S \star \hat S_\text{air} \star \hat S.   
\end{equation}
Here, $\hat S$ is the S-matrix of a single mirror and $\hat S_\text{air}$ is the scattering matrix of air layer:
\begin{equation}
\hat{S}_{\text{air}} =
\begin{bmatrix}
0 & e^{i\omega L/c} \\
e^{i\omega L/c} & 0
\end{bmatrix}.
\end{equation}

The straightforward calculation of the Redheffer star product gives the total scattering matrix of the Fabry-Pérot resonator formed by two mirrors:
\begin{equation}\label{eq:s-matrix}
\hat{S}_{\text{FP}} =
\frac{1}{1 - r^2 e^{2i\omega L/c}}
\begin{bmatrix}
r (1 - e^{2i\omega L/c}) & -t^2 e^{i\omega L/c} \\
-t^2 e^{i\omega L/c} & r (1 - e^{2i\omega L/c})
\end{bmatrix}.
\end{equation}

\end{document}